\documentclass[aps,amsmath,prb,twocolumn,a4paper,showpacs]{revtex4}
\usepackage{graphicx}
\begin{document}

%%%%%%%%%%%%%%%%%%%%%%%%%%%%%%%%%%%%%%%%%%%%%%%%%%%%%%%%%%%%%%%%%%%%%%%%%%%%%%%%%%%
\newcommand{\figureheight}{8.2 cm}
\newcommand{\putfig}[2]{\begin{figure}[h]
\special{isoscale #1.bmp, \the\hsize \figureheight}
\vspace{\figureheight} \caption{#2} \label{fig:#1}
\end{figure}}

% almost universal commands for equations and references
\newcommand{\eqn}[1]{(\ref{#1})}
\newcommand{\be}{\begin{equation}}
\newcommand{\ee}{\end{equation}}
\newcommand{\bea}{\begin{eqnarray}}
\newcommand{\eea}{\end{eqnarray}}
\newcommand{\bean}{\begin{eqnarray*}}
\newcommand{\eean}{\end{eqnarray*}}
\newcommand{\nn}{\nonumber}
%%%%%%%%%%%%%%%%%%%%%%%%%%%%%%%%%%%%%%%%%%%%%%%%%%%%%%%%%%%%%%%%%%%%%%%%%%%%%%%%%%%
%\draft

\title{Magnetic field effects on low dimensional electron systems: Luttinger liquid behaviour in a Quantum Wire}
\author{S. Bellucci $^1$ and P. Onorato $^1$ $^2$ \\}
\address{
$^1$INFN, Laboratori Nazionali di Frascati,
P.O. Box 13, 00044 Frascati, Italy. \\
$^2$Dipartimento di Scienze Fisiche, Universit\`{a} di Roma Tre,
Via della Vasca Navale 84, 00146 Roma, Italy}
\date{\today}
%\maketitle %\widetext
\begin{abstract}
We discuss the effects of a strong magnetic field in Quantum
Wires. We show how the
presence of a magnetic field modifies the role played by % which %coefficients corresponding to
electron electron interaction % and the Fermi velocity
producing a strong reduction of the backward scattering
corresponding to the Coulomb repulsion. We discuss the
consequences of this and other effects of magnetic field on the
Tomonaga-Luttinger liquids and especially on their power-law
behaviour in all correlation functions.
%We analyze with some details how the presence of strong transverse magnetic fields favors the formation of a chiral Luttinger Liquid and can induce a spin polarization.

The focal point is the rescaling of all the repulsive terms of the
interaction between electrons with opposite momenta, due to the
edge localization of the electrons and to the reduction of the
length scale. Because of the same two reasons there are some
interesting effects of the magnetic field concerning the backward
scattering due to the presence of one impurity and the
corresponding conductance. As an effect of the magnetic field we
find also a spin polarization induced by a combination of
electrostatic forces and the Pauli principle, quite similar to the
one observed in large Quntum Dots.

% Some of the obtained results could be generalized to Carbon Nanotubes and other one dimensional nanostructures.
\end{abstract}

\maketitle

\section{INTRODUCTION}

%%%%%%%%%%%PROPRIETA` SPIN ORBITA

In the last 20 years progresses in semiconductor device
fabrication and carbon technology allowed the construction of
several new devices at the nanometric scale and many novel
transport phenomena have been revealed in mesoscopic
low-dimensional structures.

Molecular beam epitaxy allows one to construct interesting
two-dimensional devices in heterostructures between different thin
semiconducting layers (a strong electric field creates a two
dimensional electron gas (2DEG) at the interface) while other
techniques (such as the electron beam lithography) for the
deposition of metallic gates allow us to confine electrons in
small devices with controllable size and contact
transparency\cite{thor}.

Semiconductor Quantum Wires (QWs) are quasi one-dimensional(1D)
devices where the electron waves are in some ways analogous to
electromagnetic waves in waveguides. They are made from a 2DEG at
the interface of a $GaAs:AlGaAs$ heterojunction where a quasi one
dimensional electron gas can be formed by etching the
heterojunction into a wire of width, say, $1000 \AA$\cite{thor}.

\

The transport in QWs is connected to three different regimes, the
two ones at very low temperatures correspond to the typical single
electron tunnelling (Coulomb blockade) and to Ballistic Transport
(where the Landauer-B\"uttiker formalism applies
)\cite{BvH,landauer2,butt} while when the correlation is strong
the Tomonaga-Luttinger\cite{kf,moroz2,moroz3,egg} liquid regime
dominates.
% and the broken simple $k,-k$ symmetry gives a subband structure rather similar to the one of Single Wall Nanotubes.

Experiments with short one-dimensional (1D) conductors (QWs, narrow
ballistic channels, quantum point contacts in a 2DEG and Carbon
Nanotubes) have demonstrated\cite{vw,war} that their conductance
is quantized in integer multiples of $2e^2/h$. However, this
simple step-like form for the conductance as a function of the
Fermi energy, occurs when the transition between the wide leads
and the narrow channel is adiabatic\cite{gla}.

The ballistic transport characterizes the motion of electrons in
nanometric regions in semiconductor structures at very high
electric field when velocities are much higher than their
equilibrium thermal velocity. We suppose that ballistic electrons
are not subjected to scattering with others electrons. A general
model for near-equilibrium ballistic transport is due to the
Landauer\cite{landauer2} and B\"uttiker\cite{butt} contributions
that are condensed in the so called Landauer formula. This formula
expresses the conductance of a system at very low temperatures and
very small bias voltages in terms of the quantum mechanical
transmission coefficients. The conductance is calculated directly
from the energy spectrum by relating it to the number of forward
propagating electron modes at a given Fermi energy.

\
%%%%%%%%%%%%%%%%%%%Luttinger

Electron transport in QWs attracts considerable interest also
because of the fundamental importance of the electron-electron
(e-e) interaction in 1D systems: the e-e interaction in a 1D
system is expected to lead to the formation of a
Tomonaga-Luttinger (TL) liquid with properties very different from
those of the non-interacting Fermi gas\cite{egg,TL,TLreviwew}.
%The e-e interaction may produce important effects in the transport that would be attractive for applications.

%%%%%%%%%%%%%%%%%%%%%%%%%%%%%%%%%%%%%%%%%%%%%%%%%%%%%%%%%

{In the TL model two types of fermions – right movers and left
movers – are coupled by an interaction of strength g. The
interaction between electrons in one-dimensional metals gives
several singular properties not present in conventional (Fermi
liquid) metals: (i) a continuous momentum distribution function
$n(k)$, varying with $k$ as $|k -k_F|^\alpha$ with an
interaction-dependent exponent $\alpha$, consequence of the lack
of fermionic quasi-particles; (ii) a similar power-law behaviour
in all correlation functions; (iii) charge-spin separation: the
elementary excitations of a TL liquid are not quasi-particles,
with charge $e$ and spin $1/2$ but collective charge and spin
density fluctuations with bosonic character, i.e. so-called
spinons and holons. These spin and charge excitations propagate
with different velocities which lead to the separation of spin and
charge.

%Because of the progresses in semiconducting and carbon technology
The interest in TL liquids increased in recent years because of
several new physical realizations, including quantum Hall edge
systems \cite{kf,wen,fn}, carbon nanotubes \cite{cnts,tubes}, and
semiconductor QWs \cite{tarucha,yacoby} in particular. Most of
these experiments concentrated on the power-law behavior of the
electron tunneling.

\

%%%%%%%%%%%%%%IMPUREZZE
The low-energy behavior of Luttinger liquids is dramatically
affected by impurities and Carbon nanotubes enable
experimentalists nowadays to analyze systems with a single
impurity in an otherwise perfectly pure one-dimensional metal.

%Theoretically such systems are usually mapped onto an effective
%field theory.
Because of the electron-electron interaction the backscattering
amplitude generated by the impurity grows at low energy scales, so
that the impurity acts as an increasingly high barrier\cite{kf2}.
A power-law singularity of the $2k_F$ density response function in
a Luttinger liquid can confirm this behavior. As a consequence,
universal scaling behavior is expected in the low energy limit,
with exponents depending only on bulk parameters of the system,
rather than on the impurity strength. }
%%%%%%%%%%%%%%%%%%%%%%%%%%%%%%%%%%%%%%%%%%%%%%%%%%%%%%%

\

%%%%%%%%%%%%%%%%%%% Spin effects and magnetic field
{A Luttinger liquid in strong magnetic field can be dramatically
modified by spin effects.} The presence of a strong magnetic
field acting on a many electron system can induce a spin
polarization. In semiconducting devices this polarization is not
an effect of Zeeman coupling but it could be a result of the
Coulomb exchange or a consequence of a transverse electric field
always present at the interface (Rashba
effect) \cite{Rashba,governale,me}. In fact the Spin Orbit (SO)
coupling due to the electric field in the $z$ direction is
stronger than the Zeeman term of interaction connected to a
magnetic field acting on the system because of the strong
reduction of the effective electron mass ($m^*=0.068 m_0$). The
Zeeman spin splitting term is $g^*\mu_B B$ where $g^*$ is the
effective magnetic factor for electrons in this geometry (very low
in $GaAs$) and $\mu_B$ is the Bohr magneton with the bare mass. So
the mass renormalization reduces by a factor $10$ the SO coupling
and by a factor $100$ the Zeeman spin splitting.

\

%%%%%%%%%%%%%%%%%%%%%Quantum Dots
The spin behavior of an electron liquid under the effect of a
strong magnetic field was accurately studied in a different
nanometric semiconducting device, the Quantum Dots (QDs), which
are small structures (typically less than $1\mu m$ in diameter)
containing from one to a few thousand electrons.
%These mesoscopic devices are useful to study a wide range of physical phenomena: from atomic like behaviour to quantum chaos, to the
%Quantum Hall Effect (QHE)\cite{Koub}.
In QDs the electronic spins align when a strong magnetic field is
present and it is known that magnetism occurs not because of
direct magnetic forces (Zeeman coupling or SO coupling), but
rather because of a combination of electrostatic forces and the
Pauli principle as was proven in the last decade in several
experiments \cite{Tar}.

\

\

In this paper we present a study of the magnetic field dependence
of Luttinger liquids in QWs. In section II we introduce the model
for a QW under the action of a magnetic field. In section III we
discuss the effect of the magnetic field on the kinetic and
interaction coefficients as well as on the derived parameters. In
particular we analyze the effects of the magnetic field on the
critical coefficient $\alpha$ which characterizes the transport of
the Wire because it determines the power law behaviour of the
Density of the States. In section III we also discuss the effects
of one impurity by analyzing the effects of magnetic field on the
conductance. In section IV we analyze how a spin polarization
could be observed in QWs in analogy with what happens in large
QDs; we also discuss the effects on Luttinger liquid behaviour due
to the spin polarization which implies the transition from a
spinful Luttinger liquid to a spinless one.

\section{Hamiltonian and microscopic approach}
%%%%%%%%%%%%%%%%%%%%%Quantum Wire
A QW is usually defined by a parabolic confining potential along
one of the directions in the plane\cite{governale,me,MRZ}:
$V(x)=\frac{m_e}{2}\omega_d^2 x^2$. We also consider a uniform
magnetic field $B$ along the $\hat{z}$ direction which allows a
free choice in the gauge determination. We choose the gauge so
that the system has a symmetry along the $\hat{y}$ direction,
${\bf A}=(0,Bx,0)$, so that the single particle Hamiltonian is
\begin{equation}\label{hnw}
H = m_e\frac{\bf v^2}{2}+\frac{m_e\omega_d^2}{2}x^2
\end{equation}
where $m_e v_y=p_y-eBx/(m_e c)$ and $m_e v_x=p_x$.

In order to solve the Hamiltonian for QWs we introduce the
cyclotron frequency $\omega_c=\frac{eB}{mc}$ and the total
frequency $\omega_T=\sqrt{\omega_d^2+\omega_c^2}$ and point out
that $p_y=v_y+eBx/(m_e c)$ commutes with the Hamiltonian
\begin{equation}\label{hnw2}
H =
\frac{\omega_d^2}{\omega_T^2}\frac{p_y^2}{2m_e}+\frac{p_x^2}{2m_e
}+\frac{m \omega_T^2}{2}(x-x_0)^2,
\end{equation}
where $x_0=\frac{\omega_c p_y}{\omega_T^2 m_e}$. The classical
Hamilton equations give us the orbital motion in the special case
of vanishing $\dot{x}(0)$
\begin{equation}\label{eqm}
\begin{array}{ll} & x(t)=x_0+R \cos(\omega_T t) \cr
& y(t)=v_d t - \frac{\omega_c}{\omega_T} R \sin(\omega_T t)+y(0)
\end{array}
\end{equation}
where the drift velocity is $v_d=\frac{\omega_d^2 p_y}{\omega_T^2
m_e}$. We obtain $p_y=m_e(\dot{y}(0)+\omega_c x(0))$, $y(0)=0$ and
$R=x(0)-x_0$ from the boundary conditions. The two different
motions along the Wire are localized on the two different edges,
as we can argue from the introduction of $\pm p_y\rightarrow \pm
v_d$. These are also known in quantum mechanics as {\it edge
states}\cite{MRZ}.

%%%%%%%%%%%%%%%%%%%%%%%%%%%%%%%%%%%%%%%%%%%%%%%%%%%%%%%%%%%%%%%%%%%%%
%\subsection{Quantum mechanics}
%%%%%%%%%%%%%%%%%%%%%%%%%%%%%%%%%%%%%%%%%%%%%%%%%%%%%%%%%%%%%%%%%%%%%
\

The quantum mechanics approach to the single particle Hamiltonian
in eq.(\ref{hnw2}) gives two term: a quantized harmonic oscillator
and a quadratic free particle-like dispersion. This kind of
factorization does not reflect itself in the separation of the
motion along each axis because the shift in the center of
oscillations along $x$ depends on the momentum $k_y$. Therefore
each electron in the system has a definite single particle wave
function
$$\varphi_{n,k_y}(x,y)=u_{n}\left(x-x_0(k_y)\right)\frac{e^{i k_y y}}{\sqrt{2 \pi L_y}},$$
$$
u_{n}\left(x-\gamma_\omega k\right)=\frac{1}{\sigma_\omega
\sqrt{\pi} }e^{-\frac{\left(x-\gamma_\omega k\right)^2}{2
\sigma_\omega^2}}h_{n}\left(x-\gamma_\omega k\right).
$$
Here $h_{n}\left(x-x_0(k_y)\right)$ is the $n$ Hermite polynomial
shifted by $x_0(k_y)=\gamma_\omega k_y$ where
$\gamma_\omega=\frac{\omega_c \hbar}{\omega_T^2 m_e}$ and
$\sigma_\omega=\sqrt{\frac{\hbar}{m_e \omega_T }}$. In what
follows we assume $\sigma_0=\sqrt{\frac{\hbar}{m_e \omega_d }}$
(corresponding to zero magnetic field) as the characteristic
length of the system when the magnetic field vanishes.

\

Now we are ready to give a simple expression of the free electron
energy depending on both the $y$ momentum $k$ and the chosen
subband $n$:
$$
\varepsilon_{n,k}=\frac{\omega_d^2}{2m_e \omega_T^2}\hbar^2
k^2+\hbar \omega_T^2(n+\frac{1}{2})
$$
Below we limit ourselves to electrons in a single channel ($n=0$),
where $N$ electrons occupy the lowest energy levels. We obtain the
value of the Fermi wavevector as $k_F=\frac{N}{4}\delta k$ where
$\delta k=\frac{2 \pi}{L}$.

%Previously we discussed that just one coefficient usually enters
%the kinetic energy, i.e. $v_F$. In fact the
The typical Luttinger model starts from the hypothesis that the
Fermi surface consists of two Fermi points, in the neighborhood of
which the dispersion curve can be approximated by straight lines
with equations
\begin{equation}\label{lin}
\varepsilon_k\approx v_F (|k|- k_F)\equiv v_F \overline{k}.
\end{equation}
In our case we obtain a field-dependent free Fermi velocity
\begin{equation}\label{vf}
v_F=\frac{\omega_d^2}{m_e \omega_T^2}\hbar k_F\approx
\frac{\omega_d^2}{m_e \omega_c^2}\hbar k_F,
\end{equation}
where the approximation is valid for very strong fields.

We introduce different operators for the electrons belonging to
each branch: right going operators ${c}^{\dag}_{R,\overline{k},s}$
and left going ones ${c}^{\dag}_{L,\overline{k},s}$ for electrons
with $\overline{k}>0$ ($\overline{k}<0$).
%The description in terms of $R/L$ movers is equivalent to the other in term of sublattices\cite{eg}).
In terms of these operators the free and interaction Hamiltonians
can be written as
\begin{eqnarray}\label{h00}
H_0&=&v_F\sum_{\overline{k},s}\overline{k}
c^{\dag}_{R,\overline{k},s} c_{R,\overline{k},s} +
v_F\sum_{\overline{k},s}\overline{k} c^{\dag}_{L,\overline{k},s}
c_{L,\overline{k},s}\\
\label{h0} H_{int}&=& \frac{1}{L}\sum_{k,k',q,s,s'}\left(
V_{k,p}^{s,s'}(q) c^{\dag}_{k+q,s} c^{\dag}_{p-q,s'}
c_{p,s'}c_{k,s}\right) .
\end{eqnarray}
Here $c_k\equiv c_{R,k}$ if $k>0$ and $c_k\equiv c_{L,k}$ if
$k<0$, while $V_{k,p}^{s,s'}(q)$ is the Fourier transform of the
electron electron interaction.

\ The scattering processes are usually classified according to the
different electrons involved and the coupling strengths labelled
with $g$ are often taken as constants. In fact, as discussed in
detail by Solyom\cite{TLreviwew}, we can substitute
$V_{k,p}^{s,s'}(q)$ with $8$ constants. In general we should take
into account the dependence on $k,p$ and $q$, however in a model
with a bandwidth cut-off, where all momenta are restricted to a
small region near the Fermi points, the momentum dependence of the
coupling is usually neglected.

We can write $g_4^{s,s'}$ for $k$ and $p$ in the same branch and
small $q$ (transferred momentum): $g_4^{s,s'}$ corresponds to the
{\em Forward Scattering in the same branch}. We use $g_2^{s,s'}$
for the {\em Forward Scattering involving two branches} where $k$
and $p$ are opposite and $q$ is small. The {\em Backward
Scattering} ($g_1^{s,s'}$) involves electrons in opposite branches
with large transferred momentum (of order $2k_F$). We do not take
into account the effects of Umklapp scattering ($g_3^{s,s'}$).

The model described above, with linear branches and constant
interaction in momentum space is known as TL model and corresponds
to a very short range interaction (Dirac delta). The presence of a
long range interaction in a 1D electron system introduces in the
model an infrared divergence and is quite difficult to solve. We
reported the solutions for the case of Carbon Nanotubes obtained
with a Renormalization Group approach and a Dimensional Crossover
in some recent papers \cite{noi}, and in the future we will apply
that formalism to the case of a QW in the presence of magnetic
field.

Below we limit ourselves to the TL model and our main results
refer to the short range interaction, with the aim of giving a
qualitative explanation of the effects of a strong transverse
magnetic field.

\section{Luttinger liquid parameters}

\subsection{Effective parameters}

%%%%%%%%%%%%%%%%%%%%% PArametri e loro ruolo
All properties of a TL liquid can be described in terms of only
two effective parameters per degree of freedom which take over in
1D the role of the Landau parameters familiar from Fermi liquid
theory.

In particular the low-energy properties of a homogeneous 1D
electron system could be completely specified by the TL
coefficients corresponding to the interaction ($g_i^{s,\sigma}$)
and the kinetic energy ($v_F$) in the limit of ideal TL liquid.

Four TL parameters, depending on $g$ and $v_F$, characterize the
low energy properties of interacting spinful electrons moving in
one channel: the parameter $\:K_{\nu}\:$ fixes the exponents for
most of the power laws and $\:v_{\nu}\:$ is the velocity of the
long wavelength excitations: $\:\nu=\rho\:$ for the charge and
$\:\nu=\sigma\:$ for the spin. The parameters\cite{hkm} $K_{\rho}$
and $v_{\rho/\sigma}$ are easily obtained as functions of
$g_i^{s,\sigma}$ and $v_F$ by various techniques found in
textbooks\cite{TLreviwew}.
\begin{eqnarray}\label{12}
K_\nu&=&\sqrt{\frac{\pi v_F+ g^\nu_4-g^\nu_2}{\pi v_F+ g^\nu_4+g^\nu_2}}\\
v_{\nu}&=&
\sqrt{\left[{v}_F+\frac{g^\nu_{4}}{\pi}\right]^2-\left(\frac{g^\nu_{2}}{\pi}\right)^2}\\
\alpha &=&
\frac{1}{2}\left[(v_F+\frac{g^\sigma_{4}}{\pi})\frac{1}{v_\sigma}
+(v_F+\frac{g^\rho_{4}}{\pi})\frac{1}{v_\rho}-2\right]\label{3}
\end{eqnarray}
where $g^\sigma_{i}=\frac{1}{2}(g^\parallel_{i}- g^\perp_{i})$ and
$g^\rho_{i}=\frac{1}{2}(g^\parallel_{i}+ g^\perp_{i})$.
Here $\alpha$ denotes the critical exponent which characterizes many
properties of the transport behaviour of a 1D device (e.g. the
zero bias conductance as a function of $T$, which is well described by a
power law behavior $G = T^\alpha$). In this
paper we discuss microscopic estimates of the values of these
quantities in semiconductor QWs when a strong transverse magnetic
field is present.
\subsection{Coupling Constants}
At the end of the previous section we discussed how the electron
electron repulsion has a very simple and detailed representation
in the TL model\cite{TLreviwew} where the $g_i^{s,s'}$
coefficients contain all the information about the interaction. In
this subsection we want evaluate the $g$ coefficients which
characterize the Hamiltonian eq.(\ref{h0}). In order to do that,
we have to calculate the "`Fourier transform"
$V^{n,n'}_{k,k'}(q,\omega_c)$ of the effective interaction
starting from some models. This is our crucial problem and gives
us the $g$ coefficients. In this article we limit ourselves to the
one channel model ($n=n'=0$) and we introduce the magnetic field
dependent effective potentials
\begin{eqnarray}\label{intpot}
\nonumber U_{k,p,q}^{s,-s}&(y&-y',\omega_c) =
\int_{\infty}^{\infty} dx dx' U(|{\bf r}-{\bf r}'| )\\ \nonumber
&\times& u_{0}\left(x-\gamma_\omega
k\right)u_{0}\left(x-\gamma_\omega p\right)
\\ \nonumber &\times& \;u_{0}\left(x'-\gamma_\omega
(k+q)\right) u_{0}\left(x'-\gamma_\omega (p-q)\right) \;.
\end{eqnarray}
These potentials only depend on $y-y'$ and on the 1D fermion
quantum numbers.

Now we are ready to calculate the "`Fourier transform" in order to
obtain $V(q)$. A central question regards the spin effects on the
coupling strength. In the Hartree-Fock approximation the so called
direct term corresponds to $q=0$ ($V(0)$) while the exchange one
corresponds to $q=p-k$ ($V(q)$). Is evident that the electron
electron repulsion is less for electrons with the same spin
($V^{\uparrow\uparrow}_q=V(0)-V(q)$)than for electrons with
opposite spins ($V^{\uparrow\downarrow}_q=V(0)$).

\

In order to analyze the effects of the range of the interaction we
introduce a function for the electron electron potential depending
on a parameter $r$. This general interaction model, ranges from
the very short range one to the infinity long range one and
eliminates the divergence of the Coulomb repulsion
\begin{eqnarray}\label{inter}
U(x-x'&,&y-y')= \frac{\left(g_0 \sigma_0^2/\pi +g_\infty r^2\right)}{r^2}\nonumber \\
&\times&\exp\left[{-\frac{(x-x')^2+(y-y')^2}{r^2}}\right],
\end{eqnarray}
where $g_0$ and $g_\infty$ are the copuling constant corresponding
to the two different limits of $r$ ($r\rightarrow 0$ delta
function $U=g_0 \sigma_0^2 \delta(|{\bf x}-{\bf x'}|)$ and
constant interaction $r\rightarrow \infty$ $U=g_\infty$).

\

So we can calculate $V_{k,p}(q,\omega_c)$ by a simple integral
\begin{eqnarray}\label{vq}
\nonumber && V_{k,p}(q,\omega_c)= \frac{\left(g_0 \sigma_0^2/\pi
+g_\infty r^2\right)\sqrt{\pi}}{L_y \left(\sqrt{r^2 + 2
\sigma_\omega^2}\right)}
\\
&&\times \exp \left[ -\frac{q^2\,r^2}{4}
-\gamma_\omega^2\left(\frac{q^2}{2 \sigma_\omega^2} +\frac{( k - p
+ q )^2}{r^2+ 2\sigma_\omega^2}\right)\right].
\end{eqnarray}
Thus we put
$$
g(r,\omega)=\frac{\left(g_0 \sigma_0^2/\pi +g_\infty
r^2\right)\sqrt{\pi}}{L_y \left(\sqrt{r^2 + 2
\sigma_\omega^2}\right)}.
$$
Some details about the comparison between Coulomb interaction and
the model of electron electron interaction in eq.(\ref{inter}) are
given in appendix.

\

Now we are ready to calculate the coefficients in eq.(\ref{h0}).

(a) The so called forward scattering in the same branch ($g_4$)
involves electrons with $p\sim k$, {\it i.e.} $\delta k= |p-k|\ll
k_F$: for parallel spins we have $g_4^\parallel\approx V_{\delta
k}(0)-V_{\delta k}(\delta k)$ while for orthogonal spins we just
have $V(0)$
$$
V_{\delta k}(0)=g(r,\omega)\exp\left[ -\gamma_\omega^2\left(
\frac{\delta k^2}{r^2+ 2\sigma_\omega^2}\right)\right]\approx
g(r,\omega).
$$
So we can assume $g_4^\parallel\approx0$ and $g_4^\perp\approx
g(r,\omega)$.

(b) The forward scattering between opposite branches, which is
usually called $g_2$ and corresponds to $p\sim-k_F,k\sim
k_F\rightarrow \delta k\approx 2 k_F$, where a small momentum is
transferred $q\sim 0$, gives as a first approximation
$$
g_2^\parallel=g_2^\perp \approx V_{2k_F}(0)=g(r,\omega)\exp\left[
-\gamma_\omega^2\left( \frac{4 k_F^2}{r^2+
2\sigma_\omega^2}\right)\right].
$$

(c) For the backward scattering $g_1$, $p\sim-k_F,k\sim
k_F\rightarrow \delta k\approx 2 k_F$ with a large momentum
transferred $q \approx 2 k_F$, we have
$$
V_{\delta k}(0)=g(r,\omega)\exp\left[-4 k_F^2\left( \frac{r^2}{4}
+\frac{\gamma_\omega^2}{2 \sigma_\omega^2}
+\frac{4\gamma_\omega^2}{r^2+ 2\sigma_\omega^2}\right)\right].
$$

Now we can discuss the effects of a magnetic field on the
interaction terms and the kinetic coefficient.

I)A strong reduction of the Fermi velocity when the magnetic field
increases is very clear, as displayed in Fig.(1) where
$v_F(\omega_C)$ is reported.

II)The forward scattering between electrons in the same branch
increases with the magnetic field as shown in Fig.(2) ($g_4\propto
g(r,\omega_c)$). We can also observe that the long range component
of this interaction is less affected by the growing field Fig.(3).

III)The backscattering ($g_1$) is strongly reduced by the magnetic
field and it gives a smaller contribution to the physics of the
system when the magnetic field increases. Also in this case the
effect is smoothed if the interaction has long range but is
however strong.

IV)The forward scattering ($g_2$) between opposite branches has a
strong reduction if we consider the short range component while
the effect is not so clear if we take into account long range
interactions.

%%%%%%%%%%%%%%%%%%%%%%%%%%%%%%%%%%%%%%%%%%%%%%%%%%%%%%%%%%%%%%%%%%%%%%%%%%%%%%%%%%%%%%%%%%%%
\begin{figure}
\includegraphics*[width=1.15\linewidth]{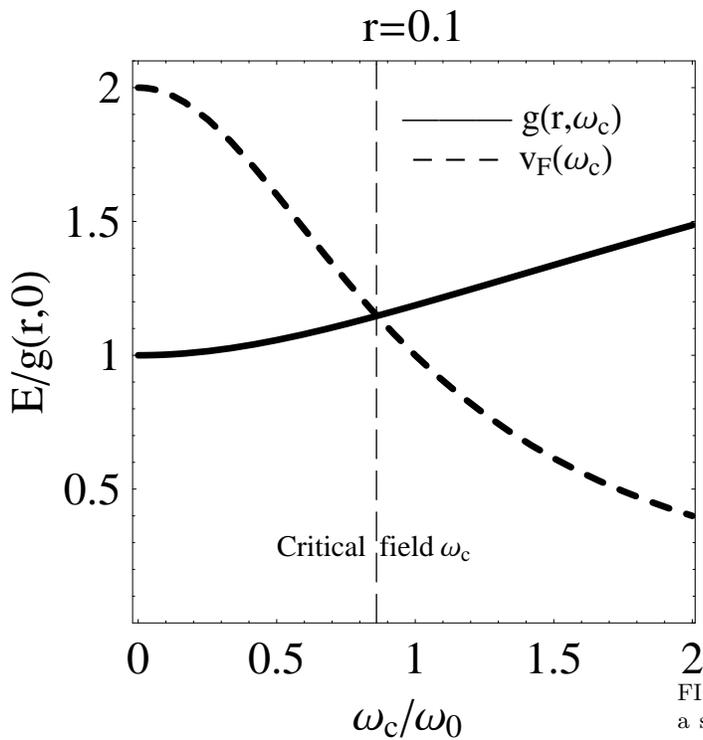}
\caption{{Graphic calculation of the critical field: the spin
transition is due to the simultaneous reduction of the Fermi
velocity and increasing of the electron electron repulsion. In the y-axis we report the
energy $E$ in units of $g(r,0)$, for a starting value
$v_F=2 g(r,0)$.
}}
\end{figure}
%%%%%%%%%%%%%%%%%%%%%%%%%%%%%%%%%%%%%%%%%%%%%%%%%%%%%%%%%%%%%%%%%%%%%%%%%%%%%%%%%%%%%%%%%%%%

%%%%%%%%%%%%%%%%%%%%%%%%%%%%%%%%%%%%%%%%%%%%%%%%%%%%%%%%%%%%%%%%%%%%%%%%%%
\begin{figure}
\includegraphics*[width=1.0\linewidth]{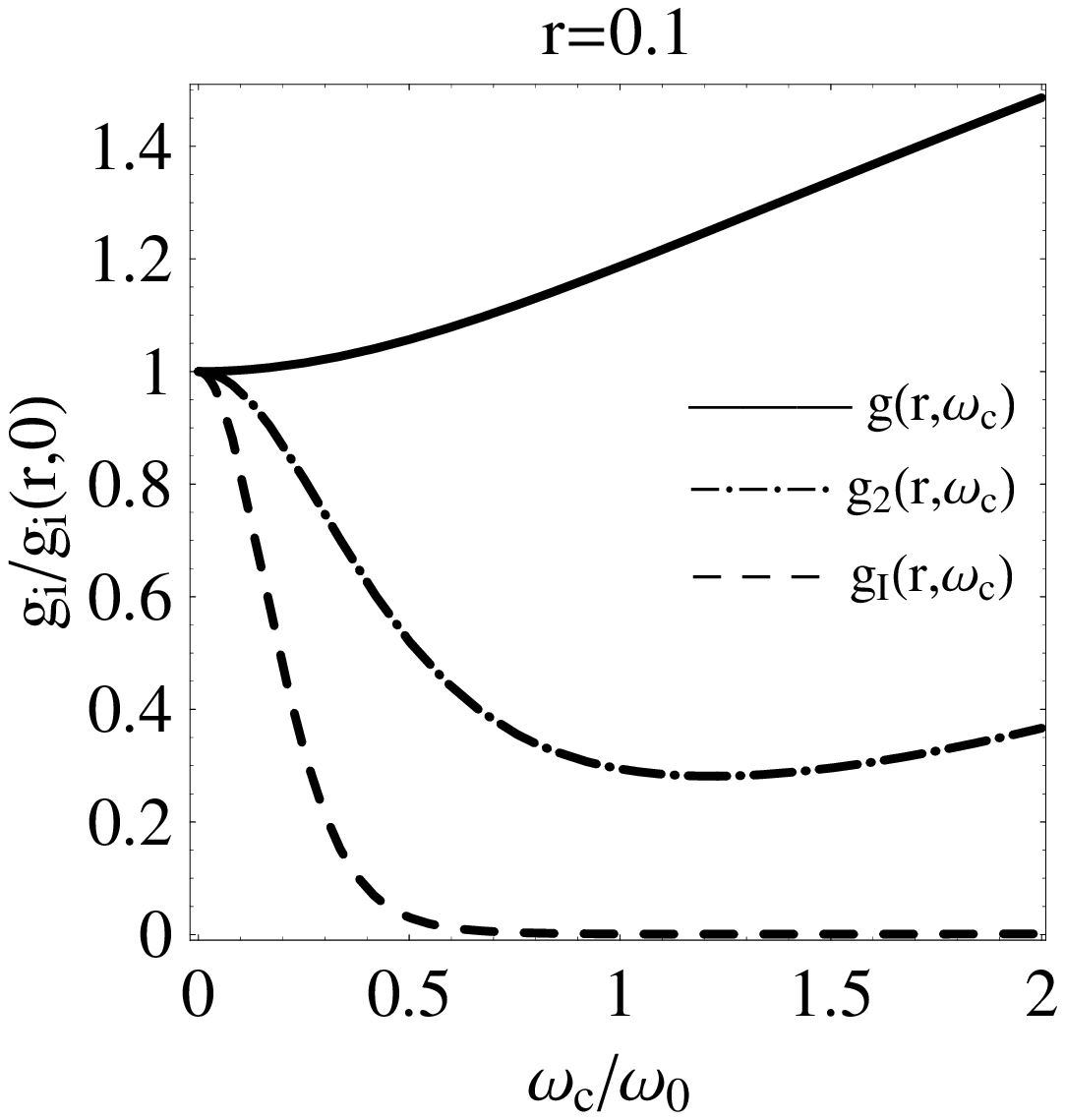}
\caption{{Scaling of the interaction with the magnetic field for a
short range of interaction $r=0.1$. Each value of $g(\omega)$ is
renormalized with respect to the value at zero magnetic field
($\omega_c=0$).}}
\end{figure}
%%%%%%%%%%%%%%%%%%%%%%%%%%%%%%%%%%%%%%%%%%%%%%%%%%%%%%%%%%%%%%%%%%%%%%%%%%

%%%%%%%%%%%%%%%%%%%%%%%%%%%%%%%%%%%%%%%%%%%%%%%%%%%%%%%%%%%%%%%%%%%%%%%%%%
\begin{figure}
\includegraphics*[width=1.10\linewidth]{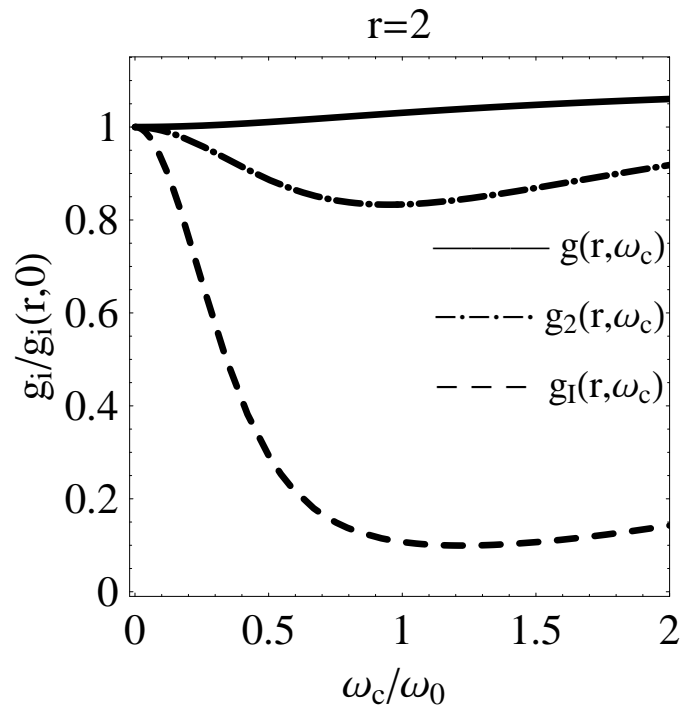}
\caption{{Scaling of the interaction with the magnetic field for a
long range of interaction $r=2$. Each value of $g(\omega)$ is
renormalized with respect to the value at zero magnetic field
($\omega_c=0$).}}
\end{figure}
%%%%%%%%%%%%%%%%%%%%%%%%%%%%%%%%%%%%%%%%%%%%%%%%%%%%%%%%%%%%%%%%%%%%%%%%%%

{ The dependence of the coupling constants $g_i$ on the magnetic
field shows a competitive effect of the current localization
against the reduction of the characteristic length
$\sigma_\omega$.

The localization of the electrons on the opposite edges of the
wire is responsible for the strong reduction of the two constant
$g_2$ and $g_1$, especially when we consider a short range
interaction. The localization is clearly seen in eq.(\ref{hnw2}),
where we introduced $x_0$ as a function of the momentum, i.e. of
the drift velocity. So we can conclude that electrons with
opposite drift velocities are localized in the opposite edges and
a reduction in the interaction has to be observed, due to the
distance between the opposite currents. Obviously this discussion
is not valid if the range of the interaction is not finite.

The effects of the localization on the coupling constants are
partially mitigated by the reduction of the length scale
$\sigma_\omega$ due to the growth of the magnetic field. The
typical length of the system reduces with the magnetic field and
increases the effective charge density of the electron liquid. As
we know, all coupling constants depend on $g(r,\omega_c)$ and
follow its behavior when the magnetic field increases. However,
just the interaction between electrons in the same branch is
really enhanced by the magnetic field if we take in account the
localization.

Because of the discussed failure of the TL model for a long
range interaction, in order to obtain the following results we
limit ourselves to the small range ($r$) case.

\subsection{The critical exponent $\alpha$}

Because of our knowledge of the coupling constants we are ready to
approach the problem of calculating the critical exponent $\alpha$
which characterizes the transport properties of the 1D electron
systems. In fact, from the transport measurements it is possible
to evaluate the Tunnel Density of States with its typical power
law dependence.}
%%%%%%%%%%%%%%%%%%%%%%%%%%%%%%%%%%%%%%%%%%%%%%%%%%%%%%%%%%%%%%%%%%%%%%%%%%%
\begin{figure}
\includegraphics*[width=1.0\linewidth]{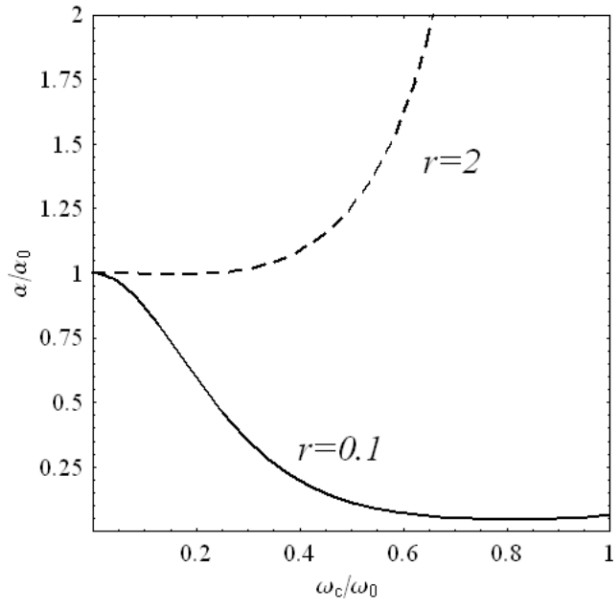}
\caption{{The critical exponent calculated following textbooks in
the limit of exactly solvable model: the range of interaction
determines either a vanishing or a divergent $\alpha$. }}
\end{figure}
%%%%%%%%%%%%%%%%%%%%%%%%%%%%%%%%%%%%%%%%%%%%%%%%%%%%%%%%%%%%%%%%%%%%%%%%%%%%
Usually the backscattering effect is not included in the models
\cite{noi} or is taken in to account as a perturbation\cite{egg},
so that here we do not consider $g_1$ more. The role of $g_2$ is
very crucial in order to determine the critical exponent
$\alpha$\cite{TL}which characterizes the linear temperature
dependence of the resistance above a crossover temperature $T_c$
in a TL liquid. When $g_2$ vanishes the so called chiral Luttinger
liquid with also $\alpha$ null appears (in that case the
spin-charge separation of excitation is present but no correlation
characterizes the Ground State).

The dependence of $\alpha$ on the magnetic field is our main
prediction and strongly depends on the effective range of
interaction. In the calculation we introduced the field-dependent
coefficients in eq.(\ref{3}) \cite{TLreviwew}. In Fig.(4) we show
the strong suppression of $\alpha$ when the magnetic field
increases for a small range potential. { Results were obtained for
values of the magnetic field $\omega_c<\omega_d$ because as we
show in the next section the behaviour of the TL liquid is
strongly modified in the high magnetic field regime. So we can
discuss the behaviour of the critical exponent just in the weak
magnetic field limit and in an intermediate one for a very short
range interaction. For a very low magnetic field we can calculate
$$
%\alpha=\frac{A}{A + v_F^0} +
% A\,\left( -\left( \frac{\frac{A}{4} - v_F^0}
% {{\left( A + v_F^0 \right) }^2} \right)
% + \frac{\frac{1}{4} - 4\,k_F}
% {A + v_F^0} \right) \,{\omega_c}^2
\alpha=\alpha_0 -\alpha_0\left(\frac{\frac{A}{4}-2 \pi v_F^0
}{(A+2 \pi v_F^0)\omega_d^2}-\frac{\frac{1}{4}-\frac{4\hbar k_F^2
}{m_e \omega_d }}{\omega_d^2} \right) \omega_c^2
$$
where $A=\frac{g_0 \sigma_0}{\sqrt{\pi}L_y}$ and
$v_F^0=\frac{\hbar k_F}{m_e}$. In the intermediate regime, where
$1/4<\omega_c/\omega_d < 3/4$, the critical exponent reduces as
follows
$$
\alpha\approx \alpha_0 e^{- \eta \frac{\omega_c}{\omega_d} }
$$
where the constant $\eta$ is a quite complicated function
depending on $A$ , $v_F^0$ and $k_F$. This behaviour is clearly
seen in Fig.(5).
%%%%%%%%%%%%%%%%%%%%%%%%%%%%%%%%%%%%%%%%%%%%%%%%%%%%%%%%%%%%%%%%%%%%%%%%%%%
\begin{figure}
\includegraphics*[width=1.2\linewidth]{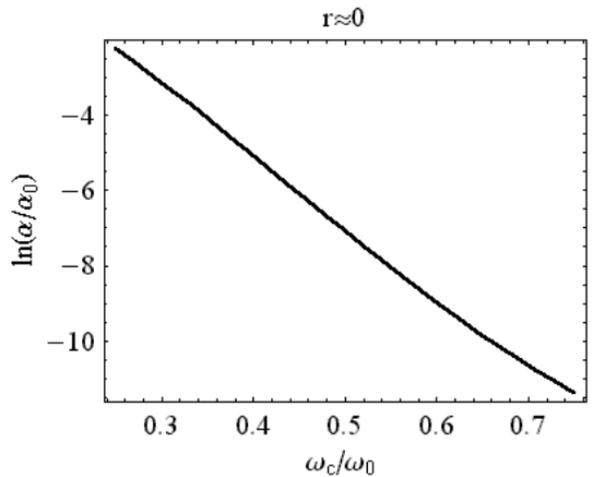}
\caption{{The critical exponent calculated the intermediate
regime, where $1/4<\omega_c/\omega_d < 3/4$, the log-log plot
shows the agreement with the exponential behaviour $ \alpha\approx
\alpha_0 e^{- \eta \frac{\omega_c}{\omega_d} } $ }}
\end{figure}
%%%%%%%%%%%%%%%%%%%%%%%%%%%%%%%%%%%%%%%%%%%%%%%%%%%%%%%%%%%%%%%%%%%%%%%%%%%%
Now we can answer the question, how the magnetic field alters the
Density of States exponents in the limit of short range
interaction: the attenuation of the forward scattering between
opposite branches due to the localization of the edge states is
responsible for the reduction of the critical exponent. This
effect dominates and characterizes the TL liquid below a value of
the magnetic field where the spin polarization crossover takes
place. At higher fields, as we discuss in the next section, the
spin polarization causes a further reduction of $\alpha$.

In Fig.(4) we also show the critical exponent calculated for a
long range interaction by using the TL model. As we discussed at
the end of sec. II, the TL model fails when it is applied to the
long range interaction because it intrinsically refers to a short
range interaction. The divergent behaviour of the exponent when
the magnetic field increases confirms the failure of the model in
treating the long range interaction.

\subsection{Impurity and Backward scattering}

Now we want to shortly discuss how the magnetic field acts on the
TL liquid when also an impurity is present in the wire. We do not
give details about calculations discussed in
refs.(\cite{kf2,fn,med}) where the problem is mapped onto an
effective field theory using bosonization and then approached by
using a Renormalization Group analysis. The usual calculations
start from a TL model with a scattering potential at the middle
point of the wire ($x=0$ and $y=0$) in which only forward
scattering is included as electron electron interaction. Two
opposite limits are usually considered: the {\em weak potential
limit} and the {\em strong potential} or {\em weak tunneling
limit}. The calculation usually starts from the transmission
probability obtained for non-interacting electrons with a Dirac
delta model for the impurity ($V(y)=V_0\delta(y)$) with
$$
|t|^2\approx \frac{1}{1+(\frac{V_0}{\hbar v_F})^2}.
$$
As discussed by Kane and Fisher\cite{kf2} the Born approximation
can be used in the small barrier limit for non-interacting
electrons. This hypothesis allows us an easy calculation of the
term which plays a central role in the scattering: $V(2k_F)$ i.e.
the Fourier transform of the potential at $2k_F$. In fact\cite{sh}
the terms which represent scattering with momentum transfer $q\ll2
k_F$ do not affect the conductance in any noticeable way, because
they do not transfer particles between $k_F$ and $-k_F$. On the
other hand, the terms which represent scattering with $|q|\simeq 2
k_F$ are expected to affect the conductance, because they change
the direction of propagation of the particles. So we introduce a
potential in order to describe the impurity localized at the
center of the wire
$$V(x,y)=U_0\frac{\sigma_0}{R_0}e^{\frac{x^2+y^2}{R_0^2}}
$$ where $R_0$ and $U_0$ represent, respectively, the range and the strength of the impurity potential.
$$
V(2k_F)=\frac{{\sqrt{2 }}\,V_0 R_0 e^{-{k_F}^2\, \left( 2\,R_0^2 +
\frac{{\gamma _\omega}^2} {{\sigma_\omega}^2} \right) }\, } {
{\sqrt{\frac{1}{2\,R_0^2} + \frac{1}{\sigma_\omega^2}}}\,
{\sigma_\omega}^2}
$$
The strong reduction of the electron backward scattering due to
the impurity depends on the magnetic field and is clearly due to
the discussed localization of the edge states. This allows us to
consider the weak potential limit, in order to proceed to the
Renormalization Group analysis. The RG equation for $V=V(2k_F)$
can be found as follows:
$$
\frac{d V}{d \ell}=\left(1-\frac{1}{2}(K_\rho+K_\sigma)
\right)V=(1-g_K)V
$$
where $E=E_0 e^{-\ell}$ is the renormalized cut-off and $E_0$ is
the original one. In absence of magnetic field we can conclude
that in our case, where $g_K<1$ corresponding to repulsive
electron electron interaction, $V(\ell)$ scales to infinity. Thus
at very low temperature ($T=0$) we have a perfect reflection.
However we can write a formula for the weak potential limit which
gives the conductance as a function of $V_0$, $g_K$ and the
temperature $T$ if the temperature is $T\gg0$
\begin{equation}\label{cond}
G_{WL}=\frac{e^2}{h}\left(1- c_0 V_0^2 T^{2 g_K- 2}+... \right)
\end{equation}
where $...$ represent higher orders in $V$ and $T$ as in
ref.\cite{fn}.
%%%%%%%%%%%%%%%%%%%%%%%%%%%%%%%%%%%%%%%%%%%%%%%%%%%%%%%%%%%%%%%%%%%%%%%%%%%
\begin{figure}
\includegraphics*[width=1.0\linewidth]{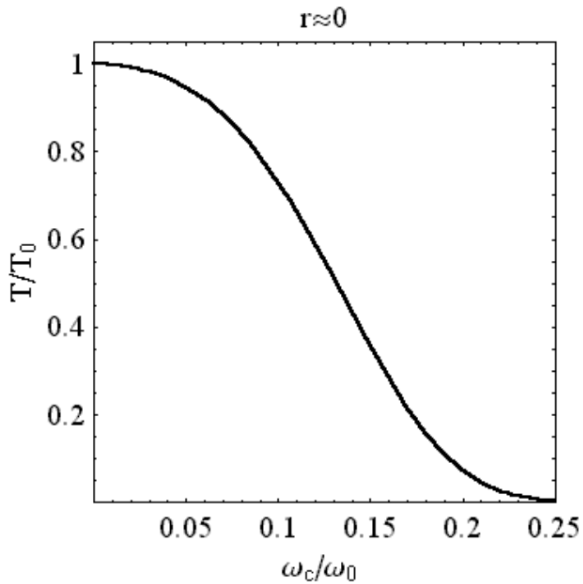}
\caption{{ The strong reduction in the value of the threshold temperature $T$, due to the
reduction of the single particle backscattering and to the
rescaling of $g_K$. The values are related to the value $T_0$,
corresponding to  the threshold temperature at zero magnetic field.
}}
\end{figure}
%%%%%%%%%%%%%%%%%%%%%%%%%%%%%%%%%%%%%%%%%%%%%%%%%%%%%%%%%%%%%%%%%%%%%%%%%%%%
As we know from the previous discussion the magnetic field acts on
eq.(\ref{cond}) by modifying both $g_k$ and $V_0$. The formula
eq.(\ref{cond}) rescales the conductance to $0$ (total reflection)
for $g_K<1$ and temperature below a threshold temperature. In the
limit of validity of the eq.(\ref{cond}) we define the threshold
temperature as the one for which the conductance vanishes
$$
T_s\approx (c_0 V_0^2)^{-\frac{1}{{2 g_K- 2}}}.
$$
As we show in Fig.(6) we obtain a strong reduction of the value of
$T_s$, due to the reduction of the single particle backscattering
and to the rescaling of $g_K$ which approaches the value $1$
corresponding to the marginal case of free electrons. }

\section{Magnetic induced phase transition: the spin polarization}

In this section we want to discuss what happens at very strong
magnetic field when a spin transition takes place in a low
dimensional electron liquid.

The spin behaviour of an electron liquid in the presence of a
growing transverse magnetic field was studied with some details in
QDs. In 1996 {Klein et al.}\cite{Klprl,Klprb,Klkram} measured the
position of the conductance peaks as a function of the magnetic
field in a large QD in the Coulomb Blockade regime\cite{nota1}.
The positions  of the peaks, due to a single electron
tunneling, allowed them to measure the ground state (GS)
energy of a many electron QD. The growth of the magnetic field
yields a crossing between energy levels so that also the GS has a
different spin polarization.  From  the measurements {Klein et
al.} could deduce that when the magnetic field is above a
threshold value, the spins flip one by one and the orbital
momentum increases.

This phenomenon has a quite general explanation in the
Hartree-Fock approximation and gives a very interesting phases
succession by increasing magnetic field. In general we can
calculate the {\em spin transition field} corresponding to the
first spin flip in the dot, and a second "critical field"
corresponding to the flip of the last spin.

\

Below we show that the magnetic field induced spin
polarization takes place also in QWs and discuss the
theoretical explanation in the general case. The physical mechanism
which induces the transitions is very simple: the kinetic energy,
proportional to the Fermi velocity, is strongly reduced by the
magnetic field while the electron electron repulsion
 is strongly enhanced by the growing field, especially the repulsion
between electrons with opposite spins (this is due to the Hund's
rule).

\ For any model with constant interaction we can find a general
condition for the spin flip and we obtain that the electron spins
flip all at the same critical field. The discrepancy between the
observed data and the prediction of this model will be better
discussed in a following article and is due to the failure of the
constant interaction model especially for the dot.
{ The condition in order to allow a spin flip is
$$
v_F(\omega_c)\delta k=g^\perp_4(\omega_c)-g^\parallel_4(\omega_c).
$$
We explain in details the case of the electrons near the Fermi
surface and the opposite one of the electrons in the bottom of the
subband at $k=0$, as we show in Fig.(7).
%%%%%%%%%%%%%%%%%%%%%%%%%%%%%%%%%%%%%%%%%%%%%%%%%%%%%%%%%%%%%%%%%%%%%%%%%%%%%%%%%%%%%%%%%%%%
\begin{figure}
\includegraphics*[width=0.95\linewidth]{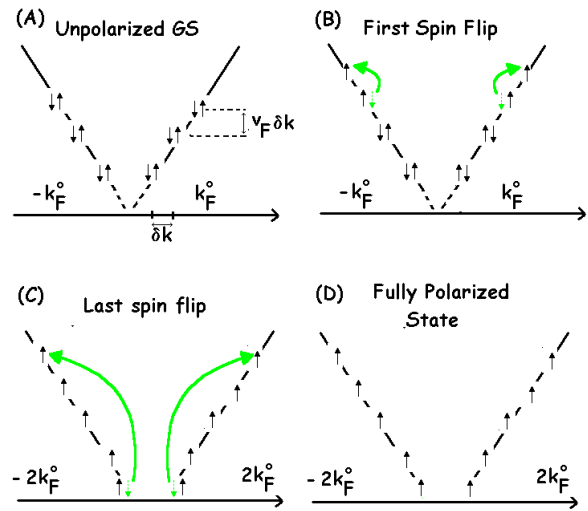}
\caption{{A) Unpolarized state with each state doubly occupied. B)
Electrons at the Fermi surface flip their spins by jumping from
the $k_F= \pm N \delta k/4$ doubly occupied to the nearest empty
levels $\pm(k_F+\delta k)$: First Splin Flip. C) Electrons at the
bottom of the subband $k=0$ flip their spins by jumping to the
first empty state: Last Spin Flip. D) The fully polarized state.
}}
\end{figure}
%%%%%%%%%%%%%%%%%%%%%%%%%%%%%%%%%%%%%%%%%%%%%%%%%%%%%%%%%%%%%%%%%%%%%%%%%%%%%%%%%%%%%%%%%%%%
}%%%%%%%%%%%%%%%%%%%%%%%%%%%
%%%%%%%%%%%%%%%%%%%%%%%%%%
%%%%%%%%%%%%%%%%%%%%%%%%%%%%%%%%%%%%%%%%%%%%%%%%%%%%%%%%%%%%%%%%%%%%%%%%%%%%%%%%%%%%%%%%%%%%
\begin{figure}
\includegraphics*[width=1.0\linewidth]{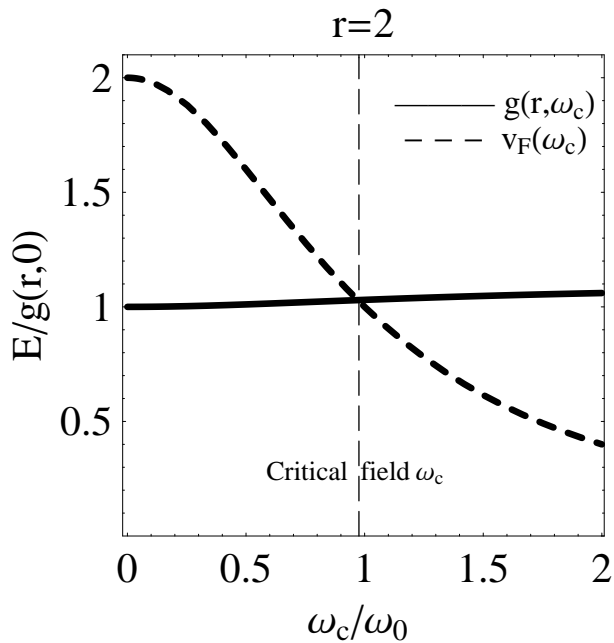}
\caption{{Graphic calculation of the critical field for a long
range interaction: the spin transition is due to the simultaneous
reduction of the Fermi velocity and increasing of the electron
electron repulsion.}}
\end{figure}
%%%%%%%%%%%%%%%%%%%%%%%%%%%%%%%%%%%%%%%%%%%%%%%%%%%%%%%%%%%%%%%%%%%%%%%%%%%%%%%%%%%%%%%%%%%%
The two electrons at the Fermi surface can flip their spins only
by jumping from the $k_F= \pm N \delta k/4$ doubly occupied to the
nearest empty levels $\pm(k_F+\delta k)$. This transition is
energetically provided if the growth in kinetic energy $2 v_F
\delta k$ is equal to the reduction in interaction energy with one
half of the remaining $(N-2)$ having spin up and the other half
with spin down ($2(g^\perp_4(\omega_c)-g^\parallel_4(\omega_c))$).
In the same way we can discuss what happens for two electrons
which jump from the bottom of the subband $k=0$ to the first empty
state, which now has all the levels singly occupied, so that the
difference in kinetic energy is $2 v_F (\frac{N}{2}-1) \delta k$,
while the gain in interaction energy is given by the spin flip
$(N-2)(g^\perp_4(\omega_c)-g^\parallel_4(\omega_c))$. Thus, we can
conclude that all the spins flip at the same critical value of the
magnetic field, provided we can consider the interaction as a
constant.

This very simple explanation fails if we assume non linear
subbands or a long range interaction. In a future article, we will
discuss this mechanism in more detail by taking in account
also the long range interaction. Here we just want to suggest
that the spin polarization takes place and it has some effects on
the interaction parameter: $g^\perp_i$ disappears and the Fermi
wavevector doubles yielding a further reduction of $g_1$ and
$g_2$.
%%%%%%%%%%%%%%%%%%%%%%%%%%%%%%%%%%%%%%%%%%%%%%%%%%%%%%%%%%%%%%%%%%%%%%%%%%%%%%%%%%%%%%%%%%%%
%\putfig{criticalbl}{Graphic calculation of the critical field: the
%spin transition is due to the simultaneous reduction of the Fermi
%velocity and increasing of the electron electron repulsion.}
%%%%%%%%%%%%%%%%%%%%%%%%%%%%%%%%%%%%%%%%%%%%%%%%%%%%%%%%%%%%%%%%%%%%%%%%%%%%%%%%%%%%%%%%%%%%
We limit ourselves to show how the strong magnetic field has the
effect of reducing the Fermi velocity and increasing the electron
electron repulsion until a critical phenomenon (spin transition)
occurs. In Figs.(1) and (8) we show the critical field from the
crossing between kinetic and repulsion energy. We can conclude
that a short range of interaction gives a lower critical field
than a long range interaction. In the figure we just show the
dependence of the critical field on the magnetic field and on the
range of the interaction.
%%%%%%%%%%%%%%%%%%%%%%%%%%%%%%%%%%%%%%%%%%%%%%%%%%%%%%%%%%%%%%%%%%%%%%%%%%%%%%%%%%%%%%%%%%%%
\begin{figure}
\includegraphics*[width=1.0\linewidth]{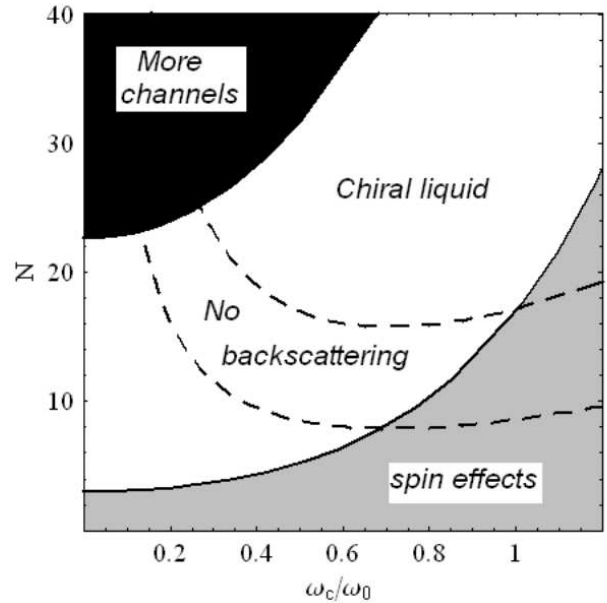}
\caption{{A qualitative phase diagram in the plane magnetic
field-number of electrons. We show that increasing the magnetic
field we have the standard case, the backscattering suppression
zone, the chiral liquid and the spin polarization. {The Number of
electrons in the vertical axis in this case is a free choice of
the authors, in general it has to be coherent with the $L$
longitudinal dimension of the wire. }} }
\end{figure}
%%%%%%%%%%%%%%%%%%%%%%%%%%%%%%%%%%%%%%%%%%%%%%%%%%%%%%%%%%%%%%%%%%%%%%%%%%%%%%%%%%%%%%%%%%%%
{ In Fig.(9) we report the different regimes in the magnetic
field-number of electrons plane, showing the various behaviours of
the electron system in the wire when the magnetic field increases.
This diagram is done in analogy with the ones obtained
experimentally and theoretically for large Quantum Dots.

In a future article we will also discuss in some details the
chiral Luttinger liquid in a ferromagnetic state analogous to a
spinless system. There, it could be interesting to analyze the
spin and charge excitations, in analogy with the Quantum Dots and
the Quantum Hall Ferromagnets. }

\section{Conclusions}
{In this work we have analyzed some properties of a Quantum Wire
when a strong magnetic field is present. Most of the effects of
the magnetic field are due to the rescaling of the electron
electron interaction and the Fermi velocity. While the Fermi
velocity always decreases with the magnetic field, the repulsive
interaction between electrons suffers the competitive actions of
the edge state localization and the characteristic length
reduction.

We have described how the growth of a transverse magnetic field
modifies the transport properties of a Quantum Wire under the
hypothesis that the latter behaves as a $1$ dimensional electron
system. In this case, i.e. the one of Luttinger liquids, the
tunneling transport properties are due to the large value of the
critical exponent $\alpha$.

At a low magnetic field the usual Luttinger liquid behavior is
predicted with some effects due to backscattering. When the
magnetic field increases there is a strong reduction of
backscattering, while for very high fields also forward scattering
between opposite branches vanishes and a chiral Luttinger liquid
appears. During the growth of the magnetic field the critical
exponent is strongly reduced. A further rise of the field can
cause the spin polarization which takes place as in a large QD,
i.e. it does not depend on Zeeman or spin orbit effects but is due
to the combined effect of the interaction and the magnetic field.

We also have discussed how the presence of one impurity can affect
the conductance in the wire. The backward scattering reduction and
the rescaling of the electron electron interaction could favor the
weak potential limit (strong tunneling) by raising the temperature
at which the wire becomes a perfect insulator ($G=0$).

In the future we wish to analyze further the possible extension of
this formalism to the study of Carbon Nanotubes and discuss with
more detail the properties of a Luttinger liquid in a fully
polarized state. }

%section{acknowledgments}

\acknowledgments

\noindent This work is partly supported by the Italian Research
Ministry MIUR, National Interest Program under grant COFIN
2002022534.
\appendix

\section{Comparison between Coulomb interaction and model}

In this appendix we discuss the difference between the Fourier
transforms of the Coulomb interaction and of the model
(eq.(\ref{intpot})).

We point out that the potential in the form of eq.(\ref{intpot})
allows exact integration, in order to obtain the Fourier
transform, and it regularizes the divergence appearing in the
Coulomb potential. This function could be optimized varying the
range parameter $r$, so that it could be quite similar to the
Coulomb potential. The function $g(\delta k,\omega_c)$ is plotted
for a very short range ($r=0.1$) and for various values of
magnetic field in Fig.(10).
%%%%%%%%%%%%%%%%%%%%%%%%%%%%%%%%%%%%%%%%%%%%%
\begin{figure}
\includegraphics*[width=1.1\linewidth]{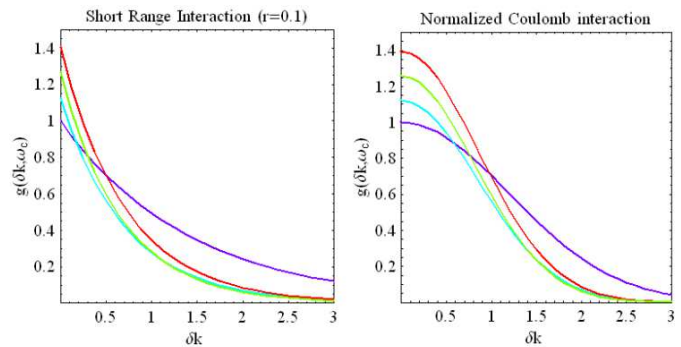}
\caption{a)On the left, the Fourier transform for various values
of magnetic field of the model of interaction with a short range
(0.1). b)On the right, the Coulomb potential case obtained with a
numerical calculation and normalized with respect to the value
obtained at zero magnetic field ($g(0,0)=1$). A comparison shows a
good agreement and an analogous magnetic field dependence.}
\end{figure}
%%%%%%%%%%%%%%%%%%%%%%%%%%%%%%%%%%%%%%%%%%%%%%
\

The Coulomb potential is not so easy to integrate: we give its
transform obtained by a numerical integration with a sort of
regularization near the divergence, and normalized with respect to
the value obtained at zero magnetic field ($g(0,0)=1$). Also in
this case we show in Fig.(10) the Fourier transform for various
values of magnetic field.
%%%%%%%%%%%%%%%%%%%%%%%%%%%%%%%%%%%%%%%%%%%%%
%\putfig{ci}{}
%%%%%%%%%%%%%%%%%%%%%%%%%%%%%%%%%%%%%%%%%%%%%%
\

Now we can conclude that the model fits well the Coulomb
interaction, if we choose a short range parameter: Fig.(10) shows
this good agreement.

We can conclude that a magnetic field in the limit of short range
gives
$$
g(0,\omega_c)\approx g(0,0)\sqrt{\frac{\omega_T}{\omega_0}}
$$
so that we have the strongest interaction between electrons with
quite similar momentum. From Fig.(10.a) and Fig.(10.b) we can also
argue that the function $V(\delta k)$ decays very rapidly when
$\delta k$ increases for high magnetic fields.

%%%--------------------------------------------------------

%%%------------------------------------------------------------------------
\bibliographystyle{prsty} %Phys. Rev. style
\bibliography{}

\begin{thebibliography}{99}
\bibitem{nota1}
This phenomenon was observed in a transverse conductance measure
where the peaks in the conductance correspond to the chemical
potential of the N-electron system.
\end{thebibliography}

\end{document}